\documentstyle[11pt,seceq]{article}

\newcommand{\be}{\begin{equation}}
\newcommand{\ee}{\end{equation}}
\newcommand{\lb}{\label}

\newcommand{\bh}{{\bf h}}

\newcommand{\bx}{{\bf x}}

\newcommand{\bz}{{\bf z}}
\newcommand{\bA}{{\bf A}}
\newcommand{\bC}{{\bf C}}
\newcommand{\bM}{{\bf M}}

\newcommand{\bR}{{\bf R}}
\newcommand{\bV}{{\bf V}}

\newcommand{\bX}{{\bf X}}

\newcommand{\bZ}{{\bf Z}}
\newcommand{\cA}{{\cal A}}
\newcommand{\cB}{{\cal B}}
\newcommand{\cC}{{\cal C}}
\newcommand{\cP}{{\cal P}}
\newcommand{\cZ}{{\cal Z}}
\newcommand{\hL}{\hat{L}}
\newcommand{\obz}{\overline{{\bf z}}}
\newcommand{\boalpha}{{\mbox{\boldmath $\alpha$}}}

\newcommand{\bomu}{{\mbox{\boldmath $\mu$}}}

\newcommand{\bozeta}{{\mbox{\boldmath $\zeta$}}}

\newcommand{\bcZ}{{\mbox{\boldmath $\cal{Z}$}}}

\newcommand{\bzed}{{\mbox{\boldmath $0$}}}
\newcommand{\boGamma}{{\mbox{\boldmath $\Gamma$}}}

\begin{document}
\title{Fluctuation-Response Relations for Multi-time Correlations}
\author{Gregory L. Eyink\\{\em Department of Mathematics}\\
{\em University of Arizona}\\{\em Tucson, AZ 85721}}
\date{ }
\maketitle
\begin{abstract}
We show that time-correlation functions of arbitrary order for any random
variable in
a statistical dynamical system can be calculated as higher-order response
functions of the
mean history of the variable. The response is to a ``control term'' added as a
modification
to the master equation for statistical distributions. The proof of the
relations is based upon
a variational characterization of the generating functional of the
time-correlations. The same
fluctuation-response relations are preserved within moment-closures for the
statistical dynamical
system, when these are constructed via the variational Rayleigh-Ritz procedure.
For the 2-time
correlations of the moment-variables themselves, the fluctuation-response
relation is equivalent
to an ``Onsager regression hypothesis'' for the small fluctuations. For
correlations of
higher-order, there is a new effect in addition to such linear propagation of
fluctuations
present instantaneously: the dynamical generation of correlations by nonlinear
interaction
of fluctuations. In general, we discuss some physical and mathematical aspects
of the
{\it Ans\"{a}tze} required for an accurate calculation of the time
correlations. We also
comment briefly upon the computational use of these relations, which is
well-suited
for automatic differentiation tools. An example will be given of a simple
closure
for turbulent energy decay, which illustrates the numerical application of the
relations.
\end{abstract}

\newpage

\section{Introduction}

It is well-known that, in statistical equilibrium systems, there are very
useful relations between
two-time correlation functions and mean response functions \cite{Kr59,DH}. The
most well-known form
of this relation gives the 2-time correlation function in terms of a response
function of the solution
of the microscopic equation of motion to an imposed infinitesimal perturbation,
when the response is
averaged over the equilibrium ensemble. These relations are often called
``fluctuation-dissipation
relations'' but we prefer the term {\it fluctuation-response relation} (FRR) as
being more
descriptive. A similiar relation has been shown to hold arbitrarily far from
thermodynamic
equilibrium in stochastic dynamical systems described by nonlinear Langevin
equations \cite{Gr77}.
In this case, however, the response is to a forcing term added into the
Fokker-Planck equation rather
than to the dynamical equation for individual realizations. The validity of
this form of the theorem
depends upon a correct coupling of the force, which, unfortunately, requires a
knowledge of the
steady-state invariant measure. This latter fact makes the generalized theorem
quite difficult to
apply in practice.

It is the purpose of this work to prove a far-reaching generalization of the
fluctuation-response
relation. Our version of the theorem holds for any (time-dependent) Markov
process described by
a master equation for the distribution function in phase space:
\be \partial_t \cP(\bx,t) = \hL(t)\cP(\bx,t), \lb{1} \ee
We include in our discussion the limiting case of the Liouville equation for a
deterministic dynamical
system. Our theorem is more similiar to that in \cite{Gr77}, since it considers
the response to a driving
or ``control'' term added into the master equation (\ref{1}) rather than to the
equation for individual
realizations. However, in contrast to that result, the coupling of our control
field does not require
any knowledge of the steady-state measure and is quite easy to write down
explicitly. Most importantly,
{\it all multi-time correlations of any finite order are obtained as
higher-order response functions
to the same control field}. Furthermore, the statistics of the system need not
be those of thermal
equilibrium or even stationary in time. The proof of the relations is based
upon a variational
characterization of the generating functional for the time-correlation
functions, which was established
in previous work \cite{Ey1}. Here we shall give a reformulation of that result
which is of interest
in its own right, as it considerably simplifies and streamlines the analysis in
the old work.

The FRR we derive is, however, prohibitively difficult to apply when (\ref{1})
describes a
spatially-extended system with many degrees-of-freedom. In such cases the
master equation is a PDE
in a huge number of variables, far too many to permit a direct numerical
solution. The practical use
of the FRR in this context will depend upon the employment of moment-closure
approximations. As we shall
show, the FRR remains valid within the moment closures when these are
formulated variationally via
the Rayleigh-Ritz method proposed in \cite{Ey1}. We shall review here the
Rayleigh-Ritz approximation,
providing some new derivations of the old results in addition to establishing
the FRR's for the
moment closures.

The plan of this paper is as follows: in Section 2 we discuss the variational
approach to statistical
dynamics. The treatment here will be new in several points and provide
significant simplifications
of that in \cite{Ey1}. We then employ the variational apparatus to establish
the FRR's. In Section 3
we review the Rayleigh-Ritz formulation of moment closure and the FRR's in that
approximation. We also
discuss there the physical significance of the FRR's, relating that for the
2-time correlation to a
linear ``regression  hypothesis''. The closure FRR for the $(n+1)$-time
correlations, with $n>1$,
contains also an effect from the nonlinear terms in the closure, namely, the
creation of correlations
by interaction of fluctuations. Some physical and mathematical properties
required of the PDF models
employed for moment closure will finally be discussed. Section 4 concerns
numerical aspects, in particular
computationally efficient and accurate methods for computing the derivatives
required in the FRR's.
A simple example of a turbulence closure will be used to illustrate the
numerical issues. The last
Section 5 will contain our conclusions from this work. An Appendix is also
included which summarizes
the results of the closure FRR diagrammatically in terms of Feynman-type graphs
with propagators and
vertices generated from the closure.

\newpage

\section {Variational Formulation \& Fluctuation-Response Relations}

\noindent {\it (2.1) Variational Approach to Statistical Dynamics}

\noindent Suppose that $\bX(t)$ is (vector-valued) Markov process, whose
distribution $\cP(\bx,t)$
at time $t$ is governed by the {\it forward Kolmogorov equation} or {\it master
equation}
\be \partial_t \cP(\bx,t) = \hL(t)\cP(\bx,t), \lb{6n} \ee
with $\hL(t)$ the instantaneous Markov generator. The random process governed
by a stochastic
differential equation is a particular example, for which the generator is the
Fokker-Planck operator.
This includes the degenerate case of a deterministic dynamics, for which the
generator is the
Liouville operator. Observables, or random variables, $\cA(\bx,t)$ evolve under
the corresponding
{\it backward Kolmogorov equation}
\be \partial_t \cA(\bx,t) = -\hL^*(t)\cA(\bx,t), \lb{6o} \ee
in which $\hL^*(t)$ is the adjoint operator of $\hL(t)$ with respect to the
canonical bilinear
form on $L^\infty\times L^1$, i.e. $<\cA,\cP>:= \int d\bx \,\,\cA(x)\cP(\bx)$.
The backward and
forward Kolmogorov equations may be simultaneously obtained as Euler-Lagrange
equations
for stationarity of the {\it action functional}
\be \Gamma[\cA,\cP]:= \int_{t_i}^{t_f} dt\,\,<\cA(t),(\partial_t-\hL(t))\cP(t)>
\lb{6p} \ee
when varied over $\cP(t)\in L^1$ with initial condition $\cP(t_i)=\cP_0$ and
$\cA(t)\in L^\infty$ with
final condition $\cA(t_f)\equiv 1$. For details, see \cite{Ey1}.

Let $\bZ(t):=\bcZ(\bX(t),t)$ be a random variable for the system given by the
continuous
function $\bcZ(\bx,t)$. Then, the {\it cumulant generating functional}
$W_Z[\bh]$ is defined as
\be W_Z[\bh]= \log\langle \exp\left(\int_{t_i}^{t_f}
dt\,\,\bh^\top(t)\bZ(t)\right)\rangle.
\lb{6b} \ee
The $n$th-order multi-time cumulants of $\bZ(t)$ are obtained from $W_Z[\bh]$
by functional
differentiation with respect to the ``test history'' $\bh(t)$:
\be C_{i_1\cdots i_n}(t_1,...,t_n)=
   \left. {{\delta^n W_Z[\bh]}\over{\delta h_{i_1}(t_1)\cdots\delta
h_{i_n}(t_n)}}\right|_{\bh=\bzed}.
\lb{6c} \ee
It is not hard to check from its definition (\ref{6b}) that $W_Z[\bh]$ is a
convex functional of $\bh$.
The Legendre dual of this functional is defined to be the effective action of
$\bZ(t)$:
\be \Gamma_Z[\bz]= \sup_{\bh}\{ <\bh,\bz>-W_Z[\bh]\}, \lb{6d} \ee
with $<\bh,\bz>:=\int dt\,\,\bh^\top(t)\bz(t)$. It is a generating functional
of so-called
{\it irreducible correlation functions} of $\bZ(t)$:
\be \Gamma_{i_1\cdots i_n}(t_1,...,t_n)=
   \left. {{\delta^n \Gamma_Z[\bz]}\over{\delta z_{i_1}(t_1)\cdots\delta
z_{i_n}(t_n)}}\right|_{\bz=\obz}.
\lb{6e} \ee
The functional derivatives here are evalutated at the mean history
$\obz(t):=\langle \bZ(t)\rangle$.
It is not hard to check from the definition (\ref{6d}) that $\Gamma_Z[\bz]$ is
a convex, nonnegative
functional with a unique global minimum (equal to zero) at the mean history
$\bz=\obz$.

There is a useful characterization of the effective action $\Gamma_Z[\bz]$ by
means of a {\it constrained
variation} of the action $\Gamma[\cA,\cP]$, which was established in
\cite{Ey1}. In fact,
\be \Gamma_Z[\bz]= {\rm st.pt.}_{\cA,\cP} \Gamma[\cA,\cP] \lb{6q} \ee
when varied over the same classes as above, but subject to constraints of fixed
overlap
\be <\cA(t),\cP(t)>= 1 \lb{6r} \ee
and fixed expectation
\be <\cA(t),\hat{\bcZ}(t)\cP(t)>= \bz(t) \lb{6s} \ee
for all $t\in [t_i,t_f]$. Note that $\hat{\bcZ}(t)$ is used to denote the
operator (in both
$L^1$ and $L^\infty$ ) of multiplication by $\bcZ(\bx,t)$. The Euler-Lagrange
equations for
this constrained variation may be obtained by incorporating the expectation
constraint (\ref{6s})
with a Lagrange multiplier $\bh(t)$. The overlap constraint may also be imposed
with
a Lagrange multiplier $\lambda(t)$, as it was in \cite{Ey1}.

However, it turns out to be advantageous to impose (\ref{6r}) through the
definitions
\begin{eqnarray}
\cA(t) & := & 1 +\left[\cB(t)-\langle\cB(t)\rangle_t\right] \cr
    \, & := & 1 + \cC(t), \lb{6t}
\end{eqnarray}
with the final conditions $\cB(t_f)=\cC(t_f)\equiv 0$. Note that
$\langle\cB(t)\rangle_t:= <\cB(t),\cP(t)>$
is the expectation with respect to the distribution $\cP(t)$. Hence, the
overlap constraint (\ref{6r})
is satisfied when $\cB(t)$ is varied independently of $\cP(t)$. The variable
$\cC(t)$ is no longer
independent of $\cP(t)$, but must satisfy the orthogonality condition
$<\cC(t),\cP(t)>=0$. The
expectation constraint must still be implemented by the Lagrange multiplier
$\bh(t)$. In terms of $\cB(t)$
or $\cC(t)$ the latter constraint is
\begin{eqnarray}
\bz(t) & = & \langle\bcZ(t)\rangle_t
       +
\left[\langle\bcZ(t)\cB(t)\rangle_t-\langle\bcZ(t)\rangle_t
\langle\cB(t)\rangle_t\right] \cr
  \,   & = & \langle\bcZ(t)\rangle_t + \langle\bcZ(t)\cC(t)\rangle_t.  \lb{6w}
\end{eqnarray}
The Euler-Lagrange equations are obtained by varying the action
$\Gamma[\cA,\cP]$ over $\cB(t),\cP(t)$
with $\cA(t)=1 +\left[\cB(t)-\langle\cB(t)\rangle_t\right]$, incorporating the
constraint (\ref{6w})
with the multiplier $\bh(t)$. A straightforward calculation gives
\be \partial_t \cP(t) =
\hat{L}(t)\cP(t)+\bh^\top(t)[\bcZ(t)-\langle\bcZ(t)\rangle_t]\cP(t)
    \lb{6u} \ee
and
\be \partial_t \cB(t)+
\hat{L}^*(t)\cB(t)+\bh^\top(t)[\bcZ(t)\cB(t)-\langle\bcZ(t)\rangle_t\cB(t)
                     -\langle\cB(t)\rangle_t\bcZ(t)]+\bh^\top(t)\bcZ(t)= 0.
    \lb{6v} \ee
Let us introduce the new operator
\be \hat{L}_\bh(t):= \hat{L}(t) +
\bh^\top(t)[\hat{\bcZ}(t)-\langle\bcZ(t)\rangle_t]. \lb{6L} \ee
The variational equations are written in terms of this operator as
\be \partial_t \cP(t) = \hat{L}_\bh(t)\cP(t)  \lb{6uu} \ee
and
\be \partial_t \cB(t)+
\hat{L}^*_\bh(t)\cB(t)+\bh^\top(t)\bcZ(t)[1-\langle\cB(t)\rangle_t]= 0.
    \lb{6vv} \ee
Using the resulting identity ${{d}\over{dt}}\langle\cB(t)\rangle_t =
-\bh^\top(t)[1-
\langle\cB(t)\rangle_t]\langle\bcZ(t)\rangle_t$, the latter equation can be
rewritten
in terms of $\cC(t)=\cB(t)-\langle\cB(t)\rangle_t$ as
\be \partial_t \cC(t)+
\hat{L}^*_\bh(t)\cC(t)+\bh^\top(t)[\bcZ(t)-\langle\bcZ(t)\rangle_t]= 0.
    \lb{6vvv} \ee

The action functional may be expressed in terms of $\cC,\cP$ as
\be \Gamma[\cC,\cP]= \int_{t_i}^{t_f} dt\,\,<\cC(t),(\partial_t-\hL(t))\cP(t)>,
\lb{6x} \ee
using $\hL^*(t)1=0$. The effective action $\Gamma_Z[\bz]$ is then obtained by
substituting
the solutions of (\ref{6uu}),(\ref{6vvv}), when the ``control field'' $\bh(t)$
is chosen so that
(\ref{6w}) reproduces the considered history $\bz(t)$. The quantity $\bz(t)$
can be seen
to be ``controllable'' by $\bh(t)$ from Legendre duality. That is, the control
$\bh[t;\bz]$
for a specified $\bz(t)$ is obtained from the minimization of the convex
function $W_Z[\bh]
-<\bh,\bz>$. Compare equation (\ref{6d}). Gathering together all of our
previous discussion
we may state:


\noindent {\it Proposition:} The effective action of the variable $\bZ(t)$ is
obtained as
\be \Gamma_Z[\bz]= \int_{t_i}^{t_f} dt\,\,<\cC(t),(\partial_t-\hL(t))\cP(t)>,
\lb{6xx} \ee
where $\cC,\cP$ satisfy
\be \partial_t \cP(t) = \hat{L}_\bh(t)\cP(t)  \lb{6uuu} \ee
and
\be \partial_t \cC(t)+
\hat{L}^*_\bh(t)\cC(t)+\bh^\top(t)[\bcZ(t)-\langle\bcZ(t)\rangle_t]= 0
    \lb{6vvvv} \ee
with initial-final conditions
\be \cP(t_i)=\cP_0,\,\,\,\cC(t_f)=0, \lb{in-fin} \ee
and the value of the control field $\bh$ is selected to give for all $t\in
[t_i,t_f]$
\be \langle\bcZ(t)\rangle_t + \langle\bcZ(t)\cC(t)\rangle_t=\bz(t). \lb{6ww}
\ee

\noindent {\it (2.2) Fluctuation-Response Relations}

\noindent It is not accidental that the same notation $\bh(t)$ was chosen above
for the control field
as for the argument of the cumulant-generating functional $W_Z[\bh]$. In fact,
we shall prove that
\be W_Z[\bh]= \int_{t_i}^{t_f} dt\,\, \bh^\top(t)\langle\bcZ(t)\rangle_t,
\lb{6y} \ee
using just the solution $\cP(t;\bh)$ of the forward equation (\ref{6uuu}), for
the control history
$\bh(t)$ which appears as the argument of $W_Z$. The result is obtained by
simply substituting the
constraint (\ref{6ww}) into the inverse Legendre transform for $W_Z$:
\begin{eqnarray}
W_Z[\bh] & = & \int_{t_i}^{t_f}dt\,\,\bh^\top(t)\bz(t) - \Gamma_Z[\bz] \cr
      \, & = &
\int_{t_i}^{t_f}dt\,\,\left\{\bh^\top(t)\left[\langle\bcZ(t)\rangle_t
               +\langle\bcZ(t)\cC(t)\rangle_t\right]
               -<\cC(t),(\partial_t-\hL)\cP(t)>\right\} \cr
      \, & = & \int_{t_i}^{t_f}dt\,\,\left\{\bh^\top(t)\langle\bcZ(t)\rangle_t
               -<\cC(t),(\partial_t-\hL_\bh)\cP(t)>\right\} \cr
      \, & = & \int_{t_i}^{t_f} dt\,\, \bh^\top(t)\langle\bcZ(t)\rangle_t,
\lb{6yy}
\end{eqnarray}
The second term in the third line vanishes by equation (\ref{6uuu}).

The relation (\ref{6y}) is a compact presentation of the fluctuation-response
relations.
Let us define the {\it response functional} of order $n$ for the variable
$\bZ(t)$ as
\be R_{i;i_1\cdots i_n}[t;t_1,...,t_n;\bh]:=
   {{\delta^n \langle\cZ_i(t)\rangle_t}\over {\delta h_{i_1}(t_1)\cdots \delta
h_{i_n}(t_n)}}[\bh],
\lb{a1} \ee
where $\langle\bcZ(t)\rangle_t$ denotes as before the average with respect to
the solution $\cP(t;\bh)$
of (\ref{6uuu}). This functional is causal, i.e. it vanishes if $t<t_i$ for any
$i=1,...n,$ because
the average $\langle\bcZ(t)\rangle_t$ cannot have any dependence upon $\bh(t')$
for $t'>t$.
The nth-order {\it response function} is taken to be the value at $\bh=\bzed$:
\be R_{i;i_1\cdots i_n}(t;t_1,...,t_n):
    = \left. R_{i;i_1\cdots i_n}[t;t_1,...,t_n;\bh]\right|_{\bh=\bzed}. \lb{a2}
\ee
We now state the {\it fluctuation-response relations:}

\noindent {\it Proposition:} The (n+1)st-order cumulant $C_{i_1\cdots
i_{n+1}}(t_1,...,t_{n+1})$ is
determined for each integer $n\geq 0$ by
\be C_{i_1\cdots i_{n+1}}(t_1,...,t_{n+1})
= \sum_{k=1}^{n+1}\,\,R_{i_k;i_1\cdots \widehat{i_k}\cdots
i_{n+1}}(t_k;t_1,...,\widehat{t_k},...,t_{n+1}),
\lb{a3} \ee
where the hat ``$\,\widehat{\,\,\,\,\,}\,$'' denotes omission of the
corresponding expression.

\noindent Observe that only the one term with $t_k=\max_p t_p$ is actually
non-zero in the sum,
by causality of the response function.

The proof of the relations is very simple. We recall the formula (\ref{6c}) for
the cumulant in terms of the
generating functional $W_Z$ and the corresponding definition (\ref{a1}) of the
response functionals.
Taking $n+1$ functional derivatives of $W_Z$ in (\ref{6y}), one obtains
\begin{eqnarray}
{{\delta^{n+1} W_Z[\bh]}\over{\delta h_{i_1}(t_1)\cdots\delta
h_{i_{n+1}}(t_{n+1})}}
    & = &  \sum_{k=1}^{n+1}\,\,R_{i_k;i_1\cdots \widehat{i_k}\cdots i_{n+1}}
                [t_k;t_1,...,\widehat{t_k},...,t_{n+1};\bh] \cr
  \,&   & \,\,\,\,\,\,\,\,\,\,\,\,+\sum_j
          \int_{t_i}^{t_f}dt\,\,h_j(t)R_{j;i_1\cdots
i_{n+1}}[t;t_1,...,t_{n+1};\bh].
\lb{a4}
\end{eqnarray}
This formula is easy to prove by induction upon $n$. Setting then $\bh=\bzed$,
the last integral
term vanishes and one obtains (\ref{a3}).

It is important to point out that the operator $\hL_\bh$ in (\ref{6uuu}) has a
quite simple and explicit
dependence upon the control field $\bh(t)$, given in (\ref{6L}). While simple,
the coupling
does depend upon the solution $\cP(t)$ itself, through the average
$\langle\bcZ(t)\rangle_t$.
Hence, the equation (\ref{6uuu}) for $\cP(t)$ is actually quadratically
nonlinear, unlike
the original master equation. Nevertheless, this nonlinearity is exactly that
required to
preserve the normalization of the solution. The equation may be integrated
forward in time to
obtain simultaneously $\cP(t)$ and $\langle\bcZ(t)\rangle_t$ as functionals of
the control $\bh$.
The response functions can then be determined by differentiating the results.

\section{Moment-Closure Approximations}

\noindent {\it (3.1) Variational formulation of moment-closure}

\noindent The results of the previous section provide a general approach to
computation of the
multi-time cumulants. However, it is obvious that the required integration of
the modified master
equation (\ref{6uuu}) will be possible only for the simplest of models, with a
few degrees of freedom.
For spatially-extended systems with many degrees of freedom, this integration
is totally intractable.
The practical employment of the FRR's then depends upon making moment-closure
approximations.
We shall review here the variational formulation of moment-closure
approximation, following
essentially the treatment in \cite{Ey1}. However, we shall also introduce some
important
simplifications, that we comment upon as we proceed.

Moment-closure approximations to the generating functional $\Gamma_Z[\bz]$ of
irreducible multi-time
correlations of $\bZ(t)$ are obtained by means of the characterization of that
functional through the
constrained variation in (\ref{6q}). Rather than varying over all $\cA\in
L^\infty,\cP\in L^1$,
one varies only over finitely parametrized trial functions. The trial functions
are constructed
from the usual elements of a moment-closure: a set of {\it moment functions}
$M_i(\bx,t),\,\,
i=1,...,R$ and a PDF {\it Ansatz} $\cP(\bx,t;\bomu)$, which is conveniently
parametrized
by the mean values which it attributes to the moment-functions, $\bomu:=\int
d\bx\,\,
\cP(\bx,t;\bomu)\bM(\bx,t)$. The left trial function is then taken to be
\be \cB(\bx,t;\boalpha):= \sum_{i=1}^R \alpha_i M_i(\bx,t). \lb{21} \ee
Following the discussion in section 2.1, we have chosen the left trial state in
the form
(\ref{6t}), to incorporate automatically the overlap constraint (\ref{6r}). The
histories
$\boalpha(t),\bomu(t)$ are the parameters to be varied over. Substituting the
trial forms,
one obtains the reduced action
\be  \Gamma[\boalpha,\bomu]= \int_{t_i}^{t_f}
dt\,\,\boalpha^\top(t)[\dot{\bomu}(t)-\bV(\bomu(t),t)]
     \lb{22} \ee
with
\be \bV(\bomu,t):= \langle (\partial_t+\hL^*)\bM(t)\rangle_{\bomu}. \lb{23} \ee
Of course, $\langle\cdot\rangle_{\bomu}$ denotes average with respect to the
PDF {\it Ansatz}.
An unconstrained variation of (\ref{22}) recovers the standard moment-closure
equation:
$\dot{\bomu}=\bV(\bomu,t)$.

For the calculation of the action, however, there is the additional expectation
constraint (\ref{6s}).
In terms of the trial parameters, it becomes
\be \bz(t)= \bozeta(\bomu(t),t)+ \bC_Z(\bomu(t),t)\boalpha(t). \lb{24} \ee
Here,
\be \bozeta(\bomu,t):= \langle\bZ(t)\rangle_\bomu \lb{25} \ee
is the $Z$-expectation within the PDF {\it Ansatz}
and
\be \bC_Z(\bomu,t):=
\langle\bZ(t)\bM^\top(t)\rangle_\bomu-\bozeta(\bomu,t)\bomu^\top \lb{26} \ee
is the corresponding $ZM$-covariance matrix. It is remarkable that
$\bozeta(\bomu,t),\bC_Z(\bomu,t)$
are the {\it only} inputs of the PDF {\it Ansatz} actually required for the
calculation.
When the constraint (\ref{24}) is incorporated into the action functional
(\ref{22}) by
means of a Lagrange multiplier $\bh(t)$, the resulting Euler-Lagrange equations
are
\begin{eqnarray}
\dot{\bomu} & = & \bV(\bomu,t)+ \bC_Z^\top(\bomu,t)\bh(t) \cr
         \, & := & \bV_Z(\bomu,\bh,t). \lb{27}
\end{eqnarray}
and
\be \dot{\boalpha}+\left({{\partial\bV_Z}\over
{\partial\bomu}}\right)^\top(\bomu,\bh,t)\boalpha
+\left({{\partial\bozeta}\over{\partial\bomu}}\right)^\top
(\bomu,t)\bh(t)=\bzed.
    \lb{28} \ee
These are solved subject to an initial condition $\bomu(t_i)=\bomu_0$ and a
final condition
$\boalpha(t_f)=\bzed$. When the solutions of the integrations are substituted
into (\ref{22}),
there results a Rayleigh-Ritz approximation $\tilde{\Gamma}_Z[\bz]$ to the
effective action
of $\bZ(t)$. The value $\bz(t)$ of the argument is that given by the constraint
equation (\ref{24})
for the given value of the control field $\bh(t)$.

The above derivation of the moment-closure approximation
$\tilde{\Gamma}_Z[\bz]$ is equivalent
to that in \cite{Ey1} but differs in some important details. The trial states
employed
in \cite{Ey1} contained each an additional parameter, $\mu_0$ and $\alpha_0$,
with
$\overline{\bomu}=\mu_0(1,\bomu),\overline{\boalpha}=(\alpha_0,\boalpha)$.
Thus,
the trial states employed there may be written, in our present notations, as
\be \cP(\bx,t;\overline{\bomu}):=\mu_0\cP(\bx,t;\bomu) \lb{28a} \ee
and
\be \cA(\bx,t;\overline{\boalpha}):= \sum_{i=0}^R \alpha_i M_i(\bx,t). \lb{28b}
\ee
Thus, $\mu_0$ was an arbitrary normalization factor and $\alpha_0$ was the
coefficient
of the constant moment-function $M_0(\bx,t)\equiv 1$. With this pair of trial
functions,
the overlap constraint (\ref{6r}) was no longer automatically enforced and
needed to
be incorporated via a Lagrange multiplier $\lambda(t)$. The resulting
Euler-Lagrange
equations of the constrained variation for
$\overline{\bomu}(t),\overline{\boalpha}(t)$
then involved both multipliers $\bh(t)$ and $\lambda(t)$. See the equations
(3.93)-(3.95)
in \cite{Ey1}. Nevertheless, those equations are equivalent to
(\ref{27})-(\ref{28}) above.
We shall not give all the details here, but leave it as a relatively simple
exercise
for the reader to check. We only point out that one may take always
$\mu_0\equiv 1$,
without any loss of generality, by absorbing that factor into the coefficients
$\boalpha$ of
the left trial function. It then follows from the $0$-component of the equation
(3.93) in
\cite{Ey1} that the Lagrange multiplier for the overlap constraint is given
explicitly as
\be \lambda(t) =  \overline{V}_0(\bomu,\bh,t)= \bh^\top(t)\bozeta(\bomu,t).
\lb{28c} \ee
If one uses this result and also uses the constraint equation (3.95) in
\cite{Ey1} to
eliminate the variable $\alpha_0$ from the equations, then one derives from
(3.93)-(3.95)
in \cite{Ey1} identically the same equations as (\ref{27})-(\ref{28}) above.

Despite their equivalence to the variational equations in \cite{Ey1}, the new
form
in (\ref{27})-(\ref{28}) above are far more convenient. Because of the presence
of the multiplier
$\lambda(t)$ in both the forward and backward equations (3.93)-(3.94) in
\cite{Ey1}, those
equations posed---apparently---a true initial-final value problem. It was
proposed
in \cite{Ey1} to solve that boundary-value problem in time with a relaxation
method.
However, we now see that the forward equation (\ref{27}) is completely
uncoupled from
the backward equation. It may be integrated forward in time, storing the
solution $\bomu(t)$
for subsequent input into the backward equation (\ref{28}) for $\boalpha(t)$.
Efficient
numerical algorithms for doing so and then integrating the results to calculate
the approximate
action have been developed by us and will be discussed in another work.

\noindent {\it (3.2) Fluctuation-response relations in closures}

\noindent A Rayleigh-Ritz approximation $\tilde{W}_Z[\bh]$ to the {\it
cumulant-generating functional}
may be introduced by the formal relation
\be \tilde{W}_Z[\bh]+\tilde{\Gamma}_Z[\bz]= <\bh,\bz> \lb{30} \ee
It may be easily checked that this definition is equivalent to the formula
\be \tilde{W}_Z[\bh]= \int_{t_i}^{t_f}dt\,\,\bh^\top(t)\bozeta(\bomu(t),t)
\lb{29} \ee
in which $\bomu(t)$ is the solution of the forward equation (\ref{27}) for the
control history
$\bh(t)$. The derivation is the exact analogue of (\ref{6yy}). Indeed, it
follows that
\begin{eqnarray}
\tilde{W}_Z[\bh] & = & \int_{t_i}^{t_f}dt\,\,\bh^\top(t)\bz(t) -
\tilde{\Gamma}_Z[\bz] \cr
              \, & = &
\int_{t_i}^{t_f}dt\,\,\left\{\bh^\top(t)\left[\bozeta(\bomu(t),t)
                       + \bC_Z(\bomu(t),t)\boalpha(t)\right]
-\boalpha^\top(t)[\dot{\bomu}(t)-\bV(\bomu(t),t)]\right\} \cr
               \, & = &
\int_{t_i}^{t_f}dt\,\,\left\{\bh^\top(t)\bozeta(\bomu(t),t)
-\boalpha^\top(t)[\dot{\bomu}(t)-\bV_Z(\bomu(t),\bh(t),t)]\right\} \cr
              \, & = & \int_{t_i}^{t_f} dt\,\,\bh^\top(t)\bozeta(\bomu(t),t),
\lb{31}
\end{eqnarray}
where (\ref{27}) was used to eliminate the second term of the third line.

It is a very attractive feature of the Rayleigh-Ritz approximation scheme that
the resulting
functionals $\tilde{\Gamma}_Z[\bz],\tilde{W}_Z[\bh]$ remain formal Legendre
transforms of each other.
That is,
\be \bh[t;\bz]={{\delta \tilde{\Gamma}_Z}\over{\delta\bz(t)}}[\bz],\,\,\,
    \bz[t;\bh]={{\delta \tilde{W}_Z}\over{\delta\bh(t)}}[\bh]. \lb{32} \ee
This follows from the discussion in \cite{Ey1}, where it was derived by a
reduction from the
underlying variational formulation of the master equation. It is worthwhile to
record here
another derivation, which is instead based directly upon the constrained moment
equations
(\ref{27})-(\ref{28}). Among other things, this new proof carries over usefully
to
discrete approximations of the moment equations employed in numerical
computations
that do not follow directly from an underlying microscopic theory.

The derivation begins by functionally differentiating (\ref{29}) with respect
to $\bh(t)$:
\begin{eqnarray}
{{\delta \tilde{W}_Z}\over{\delta\bh(t)}}[\bh] & = & \bozeta(t)+
\int_{t_i}^{t_f}ds\,\,\left({{\delta\bozeta(s)}
\over{\delta\bh(t)}}\right)^\top\bh(s) \cr
 \, & = & \bozeta(t)+
\int_{t}^{t_f}ds\,\,\left({{\delta\bomu(s)}\over{\delta\bh(t)}}\right)^\top
\left({{\partial\bozeta}\over{\partial\bomu}}(\bomu,s)\right)^\top\bh(s).
\lb{33}
\end{eqnarray}
The abbreviation $\bozeta(t):=\bozeta(\bomu(t),t)$ was introduced and in the
second line the
chain rule was employed. Causality was also invoked to reset the lower limit of
integration.
The response matrix ${{\delta\bomu(s)}\over{\delta\bh(t)}}$ that now appears
satisfies an equation
obtained by functionally differentiating (\ref{27}) with respect to $\bh(t)$:
\be {{\delta\dot{\bomu}(s)}\over{\delta\bh(t)}} =
\bA(s){{\delta\bomu(s)}\over{\delta\bh(t)}}
                           + \bC_Z^\top(\bomu,s)\delta(s-t) \lb{34} \ee
with the abbreviation
\be \bA(s):= {{\partial\bV_Z}\over{\partial\bomu}}(\bomu(s),\bh(s),s). \lb{35}
\ee
The equation (\ref{34}) can be solved with a Greens function given by a
time-ordered exponential:
\be {{\delta\bomu(s)}\over{\delta\bh(t)}}= {\rm T}\exp\left[\int_t^s
\,\,\bA(r)\,dr\right]
    \bC_Z^\top(\bomu,t). \lb{36} \ee
The adjoint of this propagator appears in the solution of the backward equation
(\ref{28}).
That equation can be written as
\be \dot{\boalpha}(t)+\bA^\top(t)\boalpha(t)
+\left({{\partial\bozeta}\over{\partial\bomu}}\right)^\top(\bomu,t)\bh(t)=\bzed
    \lb{37} \ee
and its solution is
\be \boalpha(t)= \int_t^{t_f}ds\,\,\overline{{\rm T}}\exp\left[\int_t^s
\,\,\bA^\top(r)\,dr\right]
\left({{\partial\bozeta}\over{\partial\bomu}}\right)^\top(\bomu,s)\bh(s).
\lb{38} \ee
Here $\overline{{\rm T}}\exp\left[\cdot\right]$ denotes the anti-time-ordered
exponential.
If the solution (\ref{36}) for the response matrix is substituted into the
second term of (\ref{33}),
and then (\ref{38}) is employed, it follows that
\be {{\delta \tilde{W}_Z}\over{\delta\bh(t)}}[\bh] = \bozeta(t)+
\bC_Z(t)\boalpha(t) =\bz(t), \lb{39} \ee
using (\ref{24}). Thus, the second relation in (\ref{32}) is proved. Of course,
the dual first relation
is obtained by functionally differentiating (\ref{30}) with respect to $\bz(t)$
and using (\ref{39}).

The Rayleigh-Ritz approximate generating functional $\tilde{W}_Z[\bh]$ given in
(\ref{29})
retains most of the remarkable features of the exact generating functional
$W_Z[\bh]$.
In particular, its value may be obtained by integrating just the forward
moment-equation (\ref{27})
for the selected control field $\bh(t)$, and then substituting the solution
$\bomu(t;\bh)$
into (\ref{29}). Fluctuation-response relations follow for the approximate
cumulants in
the same way as before. Just as before, one may define the {\it approximate
response function}
of order $n$ for the variable $\bZ(t)$ as
\be \tilde{R}_{i;i_1\cdots i_n}(t;t_1,...,t_n):= \left.
{{\delta^n \zeta_i(t)}\over {\delta h_{i_1}(t_1)\cdots \delta h_{i_n}(t_n)}}
\right|_{\bh=\bzed}. \lb{40} \ee
This functional is also causal. By the same argument as before, one obtains the
{\it fluctuation-
response relations} for the Rayleigh-Ritz approximate cumulants generated from
$\tilde{W}_Z[\bh]$:
\be \tilde{C}_{i_1\cdots i_{n+1}}(t_1,...,t_{n+1})=
            \sum_{k=1}^{n+1}\,\,\tilde{R}_{i_k;i_1\cdots \widehat{i_k}\cdots
i_{n+1}}
                                (t_k;t_1,...,\widehat{t_k},...,t_{n+1}).
\lb{41} \ee
This FRR within moment-closures is a practical way to compute multi-time
correlations numerically,
as we shall see below.

\noindent {\it (3.3) Physical Interpretation of the closure FRR}

\noindent The results are easiest to interpret physically in the case $n=1$. In
that case, the FRR
deals with the 2nd-order cumulant of $\bZ(t)$ or the 2-time covariance matrix
$\bC(t,t_0):=
\langle\delta\bZ(t)\delta\bZ^\top(t_0)\rangle$ (with
$\delta\bZ(t):=\bZ(t)-\langle\bZ(t)\rangle$).
The FRR here states that
\be \tilde{\bC}(t,t_0)= \tilde{\bR}(t,t_0)+[\tilde{\bR}(t_0,t)]^\top, \lb{42}
\ee
with $\tilde{\bR}(t,t_0):=\left.{{\delta\bozeta(t)}
\over{\delta\bh(t_0)}}\right|_{\bh=\bzed},$
the response matrix. Thus, for $t>t_0$,
\begin{eqnarray}
\tilde{\bC}(t,t_0) & = &
\left.{{\delta\bozeta(t)}\over{\delta\bh(t_0)}}\right|_{\bh=\bzed} \cr
         \, & = & \left({{\partial\bozeta}\over{\partial\bomu}}\right)(\bomu,t)
\left.{{\delta\bomu(t)}\over{\delta\bh(t_0)}}\right|_{\bh=\bzed} \cr
         \, & = &
\left({{\partial\bozeta}\over{\partial\bomu}}\right)(\bomu,t)\cdot
                  {\rm T}\exp\left[\int_{t_0}^t \,\,\bA_*(s)\,ds\right]
                  \bC^\top_Z(\bomu,t_0)
\lb{43}
\end{eqnarray}
using (\ref{36}). We set $\bA_*(t):=\left.\bA(t)\right|_{\bh=\bzed}$. Recall
also
that $\bC^\top_Z(\bomu,t_0)=\langle\delta\bM(t_0)\delta\bZ^\top(t_0)\rangle$.

We see that the same result can be obtained by making two physically-motivated
approximations.
The first is the {\it slaving hypothesis}: that fluctuations of the variable
$\bZ(t)$ are
instantaneously slaved to those of the moment-variables $\bM(t)$, or
$\delta\bZ(t)\approx
\left({{\partial\bozeta}\over{\partial\bomu}}\right)(\bomu,t)\delta\bM(t)$.
Thus,
\be \langle\delta\bZ(t)\delta\bZ^\top(t_0)\rangle\approx
    \left({{\partial\bozeta}\over{\partial\bomu}}\right)(\bomu,t)
     \langle\delta\bM(t)\delta\bZ^\top(t_0)\rangle. \lb{44} \ee
The second approximation is the {\it regression hypothesis}: that fluctuations
of the
moment-variables $\bM(t)$ decay on average according to the linearized closure
equation,
$\delta\dot{\bM}(t)\approx \bA_*(t)\delta\bM(t)$. Thus,
\be \langle\delta\bM(t)\delta\bZ^\top(t_0)\rangle
    \approx {\rm T}\exp\left[\int_{t_0}^t \,\,\bA_*(s)\,ds\right]
    \langle\delta\bM(t_0)\delta\bZ^\top(t_0)\rangle. \lb{45} \ee
Together, (\ref{44}) and (\ref{45}) lead directly back to the result
(\ref{43}).

\newpage

The special case of the FRR in (\ref{42}) with $n=1$ and with $\bZ(t)$ taken to
be the moment-variable
$\bM(t)$ itself, was previously derived in \cite{Ey2}. It was already pointed
out there that
the physical interpretation of the approximate FRR was provided by the
regression hypothesis.
That result has now been generalized to the case where $\bZ(t)\neq\bM(t)$ and
the new physical
principle in the Rayleigh-Ritz approximation is the slaving hypothesis.

Of course, it is also of interest to consider the physical meaning of the
approximations
involved for $n>1$. For the next case, $n=2$, the object of interest is the
3-order cumulant
\be C_{ijk}(t_2,t_1,t_0):=\langle\delta Z_i(t_2)\delta Z_j(t_1)\delta
Z_k(t_0)\rangle. \lb{46} \ee
We consider, without loss of generality, the case $t_2>t_1>t_0$. Using the FRR
and the chain rule
twice, we obtain
\begin{eqnarray}
\tilde{C}_{ijk}(t_2,t_1,t_0)
          & = & \left.{{\delta^2 \zeta_i(t_2)}\over{\delta h_j(t_1)\delta
h_k(t_0)}}\right|_{\bh=\bzed}\cr
       \, & = & {{\partial^2
\zeta_i}\over{\partial\mu_l\partial\mu_m}}(\bomu,t_2)
                \left.{{\delta\mu_l(t_2)}\over{\delta
h_j(t_1)}}\right|_{\bh=\bzed}
                \left.{{\delta\mu_m(t_2)}\over{\delta
h_k(t_0)}}\right|_{\bh=\bzed} \cr
       \, &   & \,\,\,\,\,\,\,\,\,\,\,\,\,\,\,\,+
                {{\delta\zeta_i}\over{\partial\mu_l}}(\bomu,t_2)
                \left.{{\delta^2 \mu_l(t_2)}\over{\delta h_j(t_1)\delta
h_k(t_0)}}\right|_{\bh=\bzed}.
\lb{47}
\end{eqnarray}
We see that the slaving principle holds in a generalized sense. Now higher
order derivative terms
in the Taylor expansion of $\zeta(\bomu,t)$ appear beyond the leading one.

In order to focus on the dynamical aspects of the Rayleigh-Ritz approximation
for $n=2$, let us
consider now just the special case $\bZ(t)=\bM(t)$, so that
$\bozeta(\bomu,t)=\bomu$.
We shall show that the FRR for $n=2$ (and $\bZ(t)=\bM(t)$) implies that
\begin{eqnarray}
\tilde{C}_{ijk}(t_2,t_1,t_0)
               & = & \int_{t_1}^{t_2}dt\,\,E_{ip}(t_2,t)
                     {{\partial^2
V_p}\over{\partial\mu_q\partial\mu_r}}(\bomu,t)
                     \tilde{C}_{rj}(t,t_1)\tilde{C}_{qk}(t,t_0) \cr
       \, &   & \,\,\,\,\,\,\,\,\,\,\,\,\,\,\,\,+\,\,E_{ip}(t_2,t_1)
       {{\partial
C_{jp}}\over{\partial\mu_q}}(\bomu,t_1)\tilde{C}_{qk}(t_1,t_0). \lb{49}
\end{eqnarray}
We have introduced the propagator of the linearized closure dynamics:
\be {\bf E}(t,t'):={\rm  T}\exp\left[\int_{t'}^{t}\,\,\bA_*(s)\,ds\right].
\lb{49a} \ee
The result (\ref{49}) is obtained by taking the second functional derivative
with respect to
$\bh(t_1)$ of the 1st-order response functional
\be \tilde{\bR}[t_2,t_0;\bh]
 ={\rm  T}\exp\left[\int_{t_0}^{t_2}\,\,\bA(s)\,ds\right]\bC(\bomu,t_0)
\lb{49b} \ee
using the simple identity (a continuous ``product rule'' of functional
differentiation)
\be {{\delta}\over{\delta h_j(t_1)}}{\rm T}\exp\left[\int_{t_0}^{t_2}
\,\,\bA(t)\,dt\right]
    = \int_{t_1}^{t_2}dt\,\,{\rm T}\exp\left[\int_{t}^{t_2}
\,\,\bA(s)\,ds\right]
      {{\delta\bA(t)}\over{\delta h_j(t_1)}}\,{\rm T}\exp\left[\int_{t_0}^t
\,\,\bA(s)\,ds\right],
\lb{48} \ee
and then setting $\bh=\bzed$.

Some physical insight into the result (\ref{49}) of the closure FRR can be
obtained by
rederiving the result in a more heuristic way. Let us make a {\it nonlinear
regression
hypothesis:} that the fluctuations evolve, in general, according to the full
closure dynamics.
By a Taylor expansion to quadratic order, one then obtains
\be \delta\dot{M}_i(t)= A_{*,ij}(t)\delta M_j(t)
+ {{1}\over{2}}{{\partial^2 V_i}\over{\partial\mu_q\partial\mu_r}}(\bomu,t)
  \delta M_q(t)\delta M_r(t)+ O\left(\delta M^3\right). \lb{49c} \ee
This equation can be rewritten in integral form as
\be \delta M_i(t_2) = E_{ip}(t_2,t_1)\delta M_p(t_1) +
{{1}\over{2}}\int_{t_1}^{t_2}
    dt\,\,E_{ip}(t_2,t){{\partial^2
V_p}\over{\partial\mu_q\partial\mu_r}}(\bomu,t)
  \delta M_q(t)\delta M_r(t)+ O\left(\delta M^3\right). \lb{49d} \ee
Multiplying by $\delta M_j(t_1)\delta M_k(t_0)$ and averaging then yields the
formula
\begin{eqnarray}
\langle\delta M_i(t_2)\delta M_j(t_1)\delta M_k(t_0)\rangle
  & \approx & {{1}\over{2}}\int_{t_1}^{t_2}
    dt\,\,E_{ip}(t_2,t){{\partial^2
V_p}\over{\partial\mu_q\partial\mu_r}}(\bomu,t)
  \langle\delta M_q(t)\delta M_r(t)\delta M_j(t_1)\delta M_k(t_0)\rangle  \cr
 \, & & \,\,\,\,\,\,\,\,\,\,\,\,\,\,+
       E_{ip}(t_2,t_1)\langle \delta M_p(t_1)\delta M_j(t_1)\delta
M_k(t_0)\rangle. \lb{49e}
\end{eqnarray}
Next, one can approximate the remaining correlations by discarding all
cumulants
of higher order than third (and disconnected terms single-time in $t$). Then,
\be \langle\delta M_q(t)\delta M_r(t)\delta M_j(t_1)\delta M_k(t_0)\rangle
    \approx \tilde{C}_{qj}(t,t_1)\tilde{C}_{rk}(t,t_0)+ (q\leftrightarrow r),
\lb{49f} \ee
which relation, substituted into the first term of (\ref{49e}) yields
precisely the first term of (\ref{49}). Likewise,
\be \langle \delta M_p(t_1)\delta M_j(t_1)\delta M_k(t_0)\rangle \approx
     -C_{pl}(t_1)C_{jm}(t_1)\Gamma_{lmq}(t_1)\tilde{C}_{qk}(t_1,t_0), \lb{49g}
\ee
where $\Gamma_{lmq}(\bomu,t_1)$ is the instantaneous 3rd-order irreducible
correlation in the PDF {\it Ansatz}. Substitution into the second term of
(\ref{49e})
yields the corresponding term of (\ref{49}) if we assume that
\be {{\partial C_{jp}}\over{\partial\mu_q}}(\bomu,t_1)=
-C_{pl}(t_1)C_{jm}(t_1)\Gamma_{lmq}(t_1)
    \lb{50} \ee
instantaneously at time $t_1$.

Thus, the two terms in (\ref{49}) have quite different physical
interpretations. The first integral
term represents triple correlations dynamically generated at intermediate times
$t_2>t>t_1$
from the correlations propagating in from times $t_1$ and $t_0$, which
subsequently then relax
to time $t_2$ according to the linearized closure dynamics. The
second-derivative ${{\partial^2 V_p}
\over{\partial\mu_q\partial\mu_r}}$ can be interpreted as an ``interaction
vertex'',
due to the nonlinear terms in the closure equation, by means of which the
fluctuations interact.
The second term, on the other hand, is not produced by any interaction of the
fluctuations.
It represents a triple correlation present instantaneously in the closure PDF
{\it Ansatz} which
is simply propagated in time by the linearized dynamics. Both of these terms,
as well as the
terms for cases $n>2$, can be expressed in terms of suitable diagrams. These
involve propagators
from the linearized dynamics and vertices both from the nonlinear terms in the
dynamics
and the higher-order correlations in the instantaneous PDF {\it Ansatz}. See
the Appendix.
Of course, these are not ``bare'' Feynman diagrams, for the Rayleigh-Ritz
approximation is highly
non-perturbative and the propagators and vertices represent ``dressed'' objects
resulting
from statistical closure.

The single-time relation (\ref{50}) that we derived heuristically is, in fact,
a necessary
condition for consistency of the Rayleigh-Ritz approximation to the triple
correlation.
Only if it is true will the 3-time correlation in (\ref{49}) coincide along the
diagonal
$t_2=t_1=t_0=t$ with the value $C_{ijk}(\bomu,t)$ calculated from the
single-time
PDF {\it Ansatz} $\cP(\bx;\bomu,t)$ which is input into the Rayleigh-Ritz
calculation.
Indeed, if we {\it assume} that (\ref{50}) holds, then the second term of
(\ref{49})
can be rewritten as
\be \tilde{C}_{ijk}^{(2)}(t_2,t_1,t_0)=
-\tilde{C}_{il}(t_2,t_1)C_{jm}(t_1)\Gamma_{lmq}(t_1)\tilde{C}_{qk}(t_1,t_0),
\lb{51a} \ee
from which it is manifest that $\tilde{C}_{ijk}(t,t,t)=C_{ijk}(\bomu,t)$.
Relations such as (\ref{50}) are also very important in other contexts within
the Rayleigh-Ritz
method. For example, a relation equivalent to (\ref{50}), or
\be \Gamma_{ijk}(\bomu,t)={{\partial
\Gamma_{ij}}\over{\partial\mu_k}}(\bomu,t), \lb{51} \ee
was employed in \cite{Ey2} to prove an $H$-theorem at quadratic order.

However, the relations (\ref{50}),(\ref{51}) are not automatically true for an
arbitrary PDF {\it Ansatz}.
They might be taken as {\it definitions} of the triple-correlations within the
Rayleigh-Ritz approximation,
which, we should remember, has available to it directly from the single-time
PDF {\it Ansatz} only the
mean $\zeta(\bomu,t)=\bomu$ and the covariance $\bC_Z(\bomu,t)=\bC(\bomu,t)$.
However, the triple
correlators $C_{ijk}(\bomu,t)$ and $\Gamma_{ijk}(\bomu,t)$ are {\it symmetric}
in their indices
$i,j,k$, while the definitions through (\ref{50}) and (\ref{51}) need not have
such symmetry.
Thus, such a definition may not be consistent. Fortunately, it is possible to
construct the closure
to ensure that (\ref{50}), (\ref{51}) hold, by employing an exponential PDF
{\it Ansatz},
such as those previously developed for Boltzmann kinetic equations in transport
theory \cite{CDL}.
See also \cite{Ey2}. Within such a closure scheme the single-time irreducible
correlators are all obtained from
a generating function:
\be \Gamma_{i_1\cdots i_n}(\bomu,t)
    ={{\partial^n H}\over{\partial\mu_{i_1}\cdots\mu_{i_n}}}(\bomu,t)
      \lb{51b} \ee
In fact, $H(\bomu,t)$ is just the {\it relative entropy}. Because of
(\ref{51b}), the consistency condition (\ref{51}), as well as all higher-order
ones, can be
automatically ensured by constructing the closure via an exponential PDF {\it
Ansatz}.
Thus, this closure methodology has a special relation with the Rayleigh-Ritz
approximation scheme.
This will be discussed in detail elsewhere \cite{EL}.

\section{The FRR in Numerical Computation}

\noindent {\it (4.1) Numerical Differentiation}

\noindent We have studied some of the properties of a PDF {\it Ansatz} required
for a physically accurate
and mathematically consistent approximation of time-correlations via the FRR's.
Another important
issue is the feasibility and accuracy of the FRR's for use in numerical
computations. Except
in special circumstances, it will not be possible to employ the FRR's
analytically and numerical
solution on the computer will be required. We have seen that the FRR's give the
time-correlations
by (functionally) differentiating the solutions of the modified closure
equation (\ref{27})
with respect to the control field $\bh$. The numerical problem is to compute
the required derivatives.
It is well-known that finite-difference approximations of derivatives are
inherently numerically
unstable, because the decrease in differentiation step $\Delta\bh$ needed to
reduce truncation
error must cause the round-off error in finite-precision arithmetic to grow. If
the FRR's
are to be a useful computational tool, better numerical differentiation methods
must be devised.

Fortunately, this problem has been encountered and solved in the context of
other dynamical
problems. One of the main application areas is {\it sensitivity analysis}, in
which the sensitivity
of the solution of a dynamical equation to changes in the initial data or to
parameters in the
equation is required \cite{Cac}. By ``sensitivity'' we mean just the Jacobian
derivative matrix
of the solution vector with respect to the parameter vector (or higher-order
derivatives).
Our problem is exactly of this form, in which the ``sensitivities'' required
are those of
the solution of the modified closure equation with respect to the added control
field $\bh$.
The numerical techniques which yield accurate derivatives in sensitivity
analysis depend
upon solving additional dynamical equations for the derivatives themselves.
There are
two general techniques for doing so, depending upon the time-order of
propagating derivatives:
the ``forward mode'' and the ``reverse mode''. These two techniques have
already been
illustrated in the context of our earlier discussion.  The equation (\ref{34})
for
the response matrix is equivalent to
\be {{\delta\dot{\bomu}(t)}\over{\delta\bh(t_0)}} =
\bA(t){{\delta\bomu(t)}\over{\delta\bh(t_0)}}
    \lb{52} \ee
integrated forward in time with initial data
\be \left.{{\delta\bomu(t)}\over{\delta\bh(t_0)}}
\right|_{t=t_0}=\bC_Z^\top(\bomu,t_0). \lb{53} \ee
Substituting the result into (\ref{33}) gives the derivative ${{\delta
\tilde{W}_Z}
\over{\delta\bh(t_0)}}[\bh].$ This illustrates the forward mode. Alternatively,
one may compute
the same derivative by integrating the adjoint equation (\ref{37}) for
$\boalpha(t)$ backward in time
and then substituting the result into the formula
\be {{\delta \tilde{W}_Z}\over{\delta\bh(t_0)}}[\bh]=\bozeta(t)+
\bC_Z(t)\boalpha(t). \lb{54} \ee
This illustrates the reverse mode. Such equations may be developed for
arbitrary derivatives and,
implemented numerically, they yield accurate and stable approximations.

Not only are these approaches numerically efficient but they can also be
largely automated.
Software for ``automatic differentiation'' is now becoming widely available:
see \cite{ADIFOR}.
Such tools directly generate from source code for numerical computation of the
solution of
the dynamical equation a corresponding code for the computation of its
derivative. There
is no need to compute required input derivatives, such as $A_{ij}={{\partial
V_i}\over{\partial\mu_j}},$
by hand. Furthermore, it easy to compute ``sensitivities'' with respect to new
perturbations,
such as those corresponding to a new class of variables $\bZ(t)$ of interest,
without requiring
extensive recalculations. The method has been tested and proved successful in
application
to real-life codes for PDE's employed in fluid dynamics and elsewhere. The
availability
of such software greatly enhances the attractiveness of the FRR's as a
computational method
to calculate time-correlations.

\noindent {\it (4.2) A Simple Example}

\noindent To illustrate the computational use of the FRR, we shall consider a
simple
closure for the decay of homogeneous, isotropic turbulence governed by the
Navier-Stokes
dynamics. This closure was originally employed by Kolmogorov to predict the
mean energy decay.
It was employed within the Rayleigh-Ritz formalism in \cite{Ey3} to predict the
2-time
correlation of the energy fluctuations. It should be emphasized that this
closure
omits a physical effect which is very important in the decay of energy
fluctuations:
their relaxation by turbulent diffusion in space. To see such effects, one must
construct the closure not just for the kinetic energy at a single point (say,
the origin)
but with the kinetic energy at {\it all} space points as closure variables. In
that case,
the closure equations contain ``eddy viscosity'' terms, which  are an important
linear relaxation
mechanism for fluctuations. Such improvements have been investigated and tested
against simulation
data in \cite{Ey4}. However, the 1-moment closure is adequate for our purpose,
which is to study
the utility of the FRR for numerical computations. The main merit of the
closure is that the
Rayleigh-Ritz 2-time correlation is given analytically: see Eq.(4.4) in
\cite{Ey3}.
This provides an objective basis of comparison for numerical results. We shall
only
consider here the 2-time correlations, i.e. the case $n=1$.

The closure we consider has just one moment function, the kinetic energy $K=
{{1}\over{2}}v^2$
at a single point in space. The moment average $\mu$ is here denoted $E$. It
obeys the equation
\be \dot{E}(t)= -\Lambda_m E^p(t) \lb{54a} \ee
in which $\Lambda_m,p$ are suitable real constants. See \cite{Ey3} for details.
The single-time covariance $C(t):= \langle[\delta K(t)]^2\rangle$ is
given in the closure by
\be C(E;t)= {{2}\over{3}}E^2, \lb{54b} \ee
which follows from assuming a Gaussian 1-point velocity distribution.
Thus, the perturbed closure equation becomes here
\be \dot{E}= V(E)+ h(t)C(E) \lb{54c} \ee
in which $V(E),C(E)$ are given by (\ref{54a}),(\ref{54b}), respectively. Now
the FRR states
that the 2-time correlation  $C(t,t_0):= \langle \delta K(t)\delta
K(t_0)\rangle$ is given
in the Rayleigh-Ritz approximation by
\be \tilde{C}(t,t_0)=\left.{{\delta E(t)}\over{\delta h(t_0)}}\right|_{h=0}
\lb{34i} \ee
for $t>t_0.$

The righthand side of (\ref{34i}) has been calculated by us numerically, in two
different ways.
The first method is based upon the observation that
\be \left.{{\delta E(t)}\over{\delta h(t_0)}}\right|_{h=0}=
          \left.{{\partial E}\over{\partial h}}(t;h)\right|_{h=0}.
    \lb{34j} \ee
where $E(t;h)$ is the solution of the closure equation for the modified initial
data
\be E(t_0;h):= E_0 + h\cdot C(E_0;t_0). \lb{34k} \ee
The closure equations were numerically integrated with a 4th-order Runge-Kutta
scheme with time
step $\Delta t= 10^{-3}$, in double precision arithmetic, but with initial data
given by (\ref{34k})
for the two small values $h=\pm 10^{-6}.$ The derivative (\ref{34j}) of the
solution at later times
$t$ was then estimated by the symmetric, 2nd-order finite difference
approximation to the derivative.
The second method for calculating the functional derivative in (\ref{34i}) is
to solve the analogue
of equation (\ref{52}) with initial data $\left.{{\delta E(t)}\over{\delta
h(t_0)}}\right|_{t=t_0}
=C(E_0,t_0),$ together with the closure equation itself. These were both
integrated numerically
by the same Runge-Kutta code as before. The matrix $\bA(t)$ that appears in
(\ref{52}) (here,
a $1\times 1$ matrix) is given analytically by
\be A_*(t)= -p\Lambda_m E^{p-1}(t) \lb{54d} \ee
and it was input directly into the code. Hence, the only errors in the
functional derivative
calculated by this second method, as in the solution of the closure dynamics
for the means,
are the 4th-order truncation errors and the round-off errors.

We show in Figure 1 the correlation $C(t,t_0)$ calculated by the FRR, compared
with the analytical
Rayleigh-Ritz solution given in Eq.(4.4) of \cite{Ey3}. Both the values
calculated by the finite difference
approximation (method 1) and the adjoined equation for the Jacobian (method 2)
are shown. As may be seen,
these agree perfectly, both with each other and with the exact result. This is
not to be unexpected in our
example, considering the high-order accuracy of our approximations and the
double precision arithmetic.
In other cases, it is hard to assess {\it a priori} the accuracy of the finite
difference approximation,
for which the optimal discretization step-size $\Delta\bh_{opt}$ may depend
upon both space and time
in an unknown manner.

It was shown in \cite{Ey2} that, in general, the same 2-time correlations
provided by the FRR are also given
by a linear Langevin equation. That model may be only formal, since the
covariance of its noise
term need not be positive. In any case, the numerical use of the Langevin model
to calculate the
2-time correlations is far less efficient than the numerical use of the FRR.
Not only
must the stochastic equation be integrated for a large enough number of
realizations
$N\gg 1$, but also in each realization random number generators must be called
in each
step of the time integration. Furthermore, the individual realizations governed
by the
Langevin dynamics shall be far less smooth in space and time than averages over
the ensemble,
and thus much smaller space and time discretization steps $\Delta \bx,\Delta t$
will be required.
The computational expense of using the Langevin model is, thus, far higher than
the FRR and is
not to be recommended. The stochastic equation is only useful for conceptual
purposes. The FRR,
on the other hand, is quite efficient because it takes full advantage of the
increased
regularity and stability of statistically-averaged quantities. It is really a
``thermodynamic
approach'' to calculating the time-correlations and not a
``statistical-mechanical'' method.

\section{Conclusions}

\noindent In this paper we have reviewed and simplified the variational
approach to statistical dynamics
proposed in \cite{Ey1}. As a main new result, we derived a general
fluctuation-response relation (FRR)
for arbitrary multi-time correlations. We demonstrated that the FRR's are
preserved in a moment-closure
approximation by the Rayleigh-Ritz method. We discussed the physical
significance of the closure FRR's
in terms of various intuitive hypotheses: slaving, regression (linear and
nonlinear) of fluctuations.
We also discussed computationally efficient and accurate methods for computing
the derivatives
required in the FRR's.

Many interesting problems can be investigated with the present methods. These
include temporal multiscaling
in turbulence \cite{BBCT,VPP}, aging phenomena in glassy relaxation
\cite{CK,BCKM}, transition rate theory
in chemical kinetics \cite{SRL}, and Lagrangian statistics of advected scalar
reactants \cite{Fox}.
The FRR should also hold for quantum systems, governed by quantum Liouville or
master equations.
Rayleigh-Ritz methods could provide a tractable means to compute multi-time
statistics in quantum
field theory and in the quantum many-body problem.

\vspace{.3in}

\noindent {\bf Acknowledgements.} The author wishes to thank F. Alexander, C.
D. Levermore and
J. Restrepo for valuable conversations and suggestions which contributed to
this work.

\noindent {\bf Appendix: Diagrammatic Rules}

\noindent We sketch concisely here the diagrammatic rules which follow
from the closure FRR for the multi-time cumulants or connected
correlation functions. The derivation generalizes that for the 3-time
cumulant in subsection 3.3. It is advantageous to specify the latest
time to be $t_n$ and the earliest time $t_0$. Then, $(n+1)$-time
correlations for $t_f=t_n>t_{n-1}>\cdots >t_t>t_0=t_i$ are obtained by
succcessive functional differentiation of expression (\ref{49b}) for
$\tilde{\bR}[t_n,t_0;\bh]$ with respect to $\bh(t_1),...,\bh(t_{n-1})$,
using the ``product rule'' (\ref{48}). We recall from (\ref{35}) that
$$ \bA(t):= {{\partial\bV}\over{\partial\bomu}}(\bomu,t)+
            \bh^\top(t){{\partial\bC}\over{\partial\bomu}}(\bomu,t) $$
and employ the chain rule to calculate successive derivatives. We also
assume that the single-time irreducible correlations have the entropy
as a generating function, as in (\ref{51b}).

The terms which result may be associated to graphs. The lines in the
graphs terminating at times $t,t'$ (internal or external) are given by
the covariance function $\tilde{C}_{ij}(t,t')$. If $t=t'$, then
$\tilde{C}_{ij}(t,t)=C_{ij}(\bomu,t)$. The vertices are of two types.
For each integer $r\geq 2$ there are $(r+1)$-fold vertices of the form
$$ W_i^{j_1\cdots j_r}(s)
    :=\Gamma_{im}(s){{\partial^r
        V_m}\over{\partial\mu_{j_1}\cdots\mu_{j_r}}}(\bomu,s) $$
and
$$ -\Gamma_{j_1\cdots j_{r+1}}(t)
    = -{{\partial^{r+1} H}\over{\partial\mu_{j_1}\cdots\mu_{j_r}}}(\bomu,t), $$
where the latter is just minus the single-time, irreducible
$(r+1)$st-order correlator. The minus sign appears because of the fact
that $\bC=\boGamma^{-1}$ and thus
$$ {{\partial\bC}\over{\partial\bomu}}= -\bC
   {{\partial\boGamma}\over{\partial\bomu}}\bC. $$
One may replace the $W$-vertices with $V_m^{j_1\cdots j_r}(s):=
{{\partial^r V_m}\over{\partial\mu_{j_1}\cdots\mu_{j_r}}}(\bomu,s) $
if the propagator line $\tilde{C}_{ki}(s',s)$ entering the $i$-node of the
$W$-vertex is replaced by a linear propagator $E_{km}(s',s)$ entering
the $m$-node of the $V$-vertex.

The following rules apply:
\begin{itemize}
\item [(i)] The graphs which appear are all tree graphs with the times
  $t_n,t_{n-1},...,t_1,t_0$ terminating the external lines. The trees
  are rooted at time $t_n$ and branch up to earlier times, with the
  times non-increasing as one ascends the tree.
\item [(ii)] Each vertex must be linked to at least one of the early
  external times $t_{n-1},...,t_0$ directly by a propagator line.
\item [(iii)] The $\Gamma$-type vertices are all evaluated at an
  early external time $t_{n-1},...,t_0$ which is determined as the
  latest time reached by any branch starting upward from that vertex
  and passing only through $\Gamma$-type vertices.
\item [(iv)] The $W$-type vertices (or the $V$-type) are evaluated at
  internal times $s$ which are integrated over the largest possible
  subrange of $t_n>s>t_0$ consistent with the rule of non-increasing times
  ascending the tree.
\end{itemize}
Because of rule (ii) it is clear that there are only finitely many
graphs contributing to each $(n+1)$-time cumulant function, with
vertices of at most $(n+1)$st-order appearing. The finite sum of all
the contributions from these graphs gives the FRR result for the
cumulant function. Thus, it is clear that this graphical
representation is not a perturbation expansion into Feynman diagrams,
since the latter contains closed loops and infinitely many terms. The
propagators and vertices here are all ``dressed objects'' and the
representation is nonperturbative.

When all the external times are equal, $t_n=\cdots=t_0=t$, then the
graphical representation simplifies considerably. There are then no
$V$ or $W$-type vertices, because the integration range over each
internal time $s$ has shrunk to zero. Furthermore, all of the
$\Gamma$-type vertices are now evaluated at the same time $t$. In
fact, the resulting graphical expansion is just that of the well-known
representation of the single-time $(n+1)$st-order cumulant
$C_{i_1\cdots i_{n+1}}(t)$ as a sum over tree diagrams with single-time
irreducible correlations $\Gamma_{i_1 \cdots i_{r+1}}(t)$ as
vertices and 2nd-order correlators $C_{ij}(t)$ on the internal and
external lines. Thus, we obtain a proof for any order $(n+1)$ that, along the
diagonal $t_n=\cdots=t_0=t$ in time, $\tilde{C}_{i_1\cdots i_{n+1}}(t,...,t)
=C_{i_1\cdots i_{n+1}}(\bomu,t)$.

\newpage

\noindent {\bf Figure 1: Comparison of Two-Time Correlations}. FRR Calculations
vs. Exact Result.
           Initial energy at $t_0=0$ was set to $E_0=10$. The decay was
calculated with
           spectral parameters: $m=2,A=1,$ and Kolmogorov constant $\alpha=2$.
           Notations follow reference \cite{Ey3}.

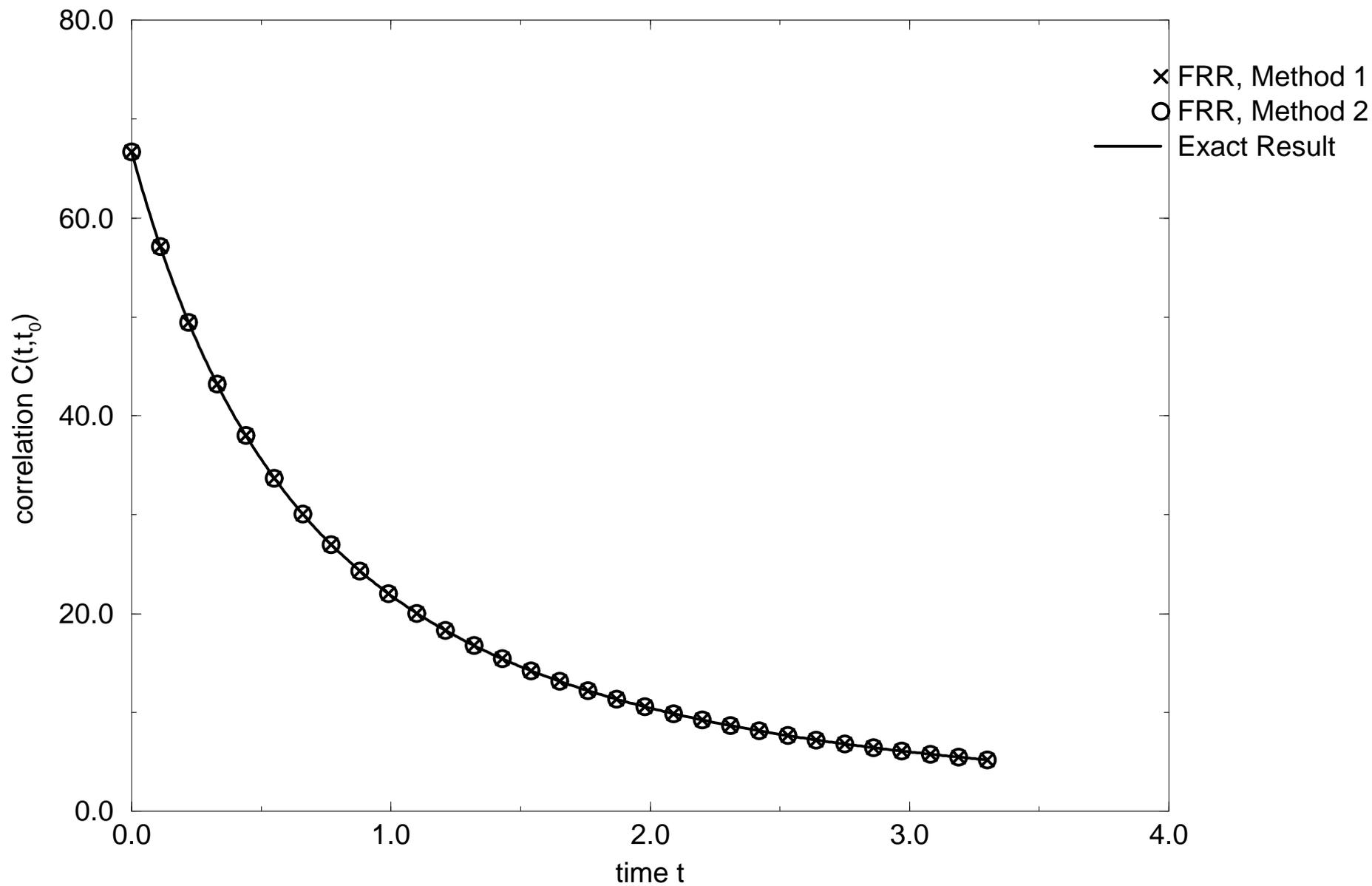
\end{document}